\documentclass[showpacs,prb,floatfix,twocolumn,amsmath]{revtex4}
\usepackage{graphicx}
\usepackage[textsize=tiny]{todonotes}

\begin{document}

\title{Domain dynamics in the multiferroic phase of MnWO$_4$}

\author{D. Niermann$^1$}
\author{C.P. Grams$^1$}
\author{M. Schalenbach$^1$}
\author{P.~Becker$^{2}$}
\author{L.~Bohat\'y$^{2}$}
\author{J.~Stein$^{1}$}
\author{M.~Braden$^{1}$}
\author{J.~Hemberger$^{1}$}
\email[Corresponding author:~]{hemberger@ph2.uni-koeln.de}

\affiliation{$^1$II.\ Physikalisches Institut, Universit\"at zu K\"oln, Z\"ulpicher Str.\ 77, D-50937 K\"oln}
\affiliation{$^2$Institut f\"ur Kristallographie, Universit\"at zu K\"oln, Greinstr.\ 6, D-50939 K\"oln, Germany}

\begin{abstract}
By using broadband linear and non-linear dielectric spectroscopy we studied the magnetoelectric dynamics in the
chiral antiferromagnet MnWO$_4$.
In the multiferroic phase the dielectric response is dominated by the dynamics of domains and domain walls which is strongly dependent on the stimulating electric field. The mean switching time reaches values in the minute range in the middle of the multiferroic temperature regime at $T\approx 10$\,K  but unexpectedly decays again on approaching the lower, first-order phase boundary at $T_{N1} \approx 7.6$\,K. 
The switchability of the ferroelectric domains denotes a pinning-induced threshold and 
 can be described considering a growth-limited scenario with an effective growth dimension of $d \approx 1.8$. 
The rise of the effective dynamical coercive field on cooling below the $T_{N2}$ is much stronger compared to the usual ferroelectrics and can be described by a power law $E_c\propto \nu^{1/2}$. The latter 
questions the feasibility of fast switching devices based on this type of material.
\end{abstract}

\date{\today}

\pacs{75.85.+t, 75.78.-n.Fg, 77.80.B-.Fm, 75.30.Mb, 75.60.Ch}

\maketitle

\subsection{Introduction}

In recent years magnetoelectric multiferroics have attracted considerable interest within the community of correlated transition-metal compounds \cite{Khomskii09,Spaldin05}. In these compounds ferroelectric order and magnetism do not only coexist in a single phase but exhibit strong coupling of the ferroic order parameters. Among other mechanisms 
the probably most established magnetoelectric coupling scenario is based on the inverse Dzyaloshinskii-Moriya (DM) interaction in partially frustrated spiral magnets \cite{Cheong07,Tokura09}. This type of magnetically driven ferroelectricity is the underlying mechanism, e.g., in the numerously studied family of multiferroic manganites such as TbMnO$_3$ \cite{Kimura03, Hemberger07} and in principle is now well understood. In these compounds a non-collinear cycloidal spin-structure is directly coupled to a ferroelectric polarization resulting from the coherent distortion of the Mn-O-Mn bonds perpendicular to the propagation vector of the spin-cycloid.  However, the manifestation of a magnetoelectrically coupled multiferroic phase and the formation of the complex, magnetoelectric order parameter raises questions concerning the corresponding dynamics.

One aspect is given by the elementary excitations within the ordered multiferroic phase, the so called electromagnons, as they were detected, e.g., in perovskite manganites in a typically sub-phononic region below tera\-hertz frequencies\cite{Pimenov06,Katsura07,Senff07}. 
%
%
Another aspect is the low frequency dielectric response originating from intrinsic or extrinsic sample inhomogeneities. In materials with a residual conductivity, which in addition may be dependent on external fields, contributions of  contacts or grain boundaries may add capacitive or even magneto-capacitive contributions, which will cover the intrinsic sample properties \cite{Lunkenheimer02,Catalan06,Niermann12,Schmidt12}. Also one may find relaxational features resulting from localized polarons at defect states as, e.g., demonstrated for the case of perovskite rare-earth manganites above and within the multiferroic phase \cite{Schrettle09}. But even though one avoids these latter contributions by choosing a suitable, i.e.\ well insulating, high purity single crystalline material as we did in this study, the formation of domains and their emergent dynamics will dominate the dielectric response of the multiferroic system \cite{Kagawa09, Meier09, Hoffmann11, Baum13}.

\begin{figure}
\centerline{\includegraphics[width=0.90\columnwidth,angle=0]{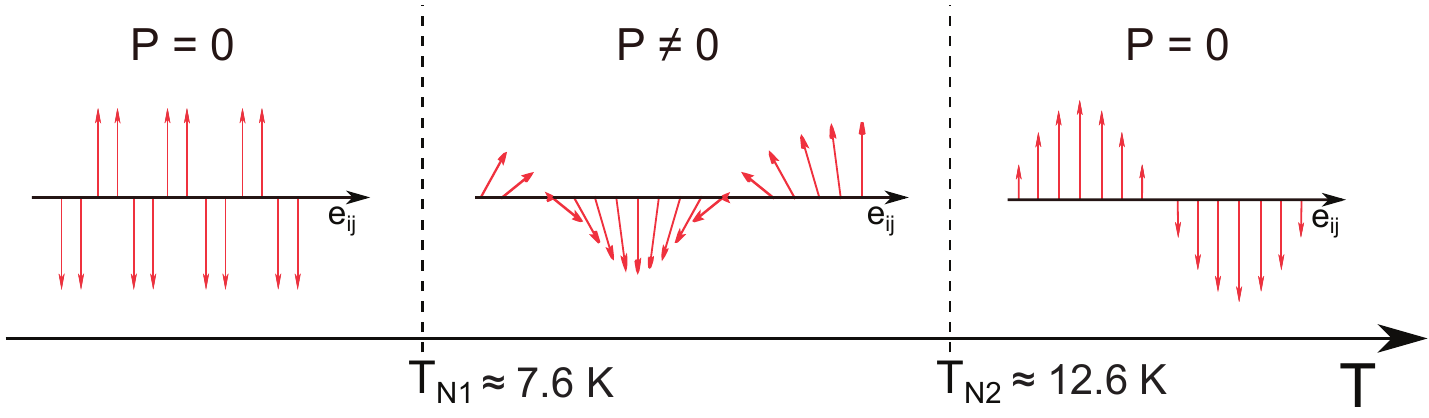}}
\caption{
Schematic sketch of the magnetically ordered phases in MnWO$_4$.[\onlinecite{Lautenschlaeger93}]
}
\label{phases}
\end{figure}

The system MnWO$_4$, the mineral name is h\"ubnerite, represents the above described class of magnetoelectric multiferroics, in which the ferroelectricity is driven by the onset of chiral spin order via the inverse DM interaction \cite{Heyer06,Taniguchi06,Arkenbout06}. Its crystal structure in the paramagnetic phase is monoclinic with space group P2/c and can be thought of as alternate stacking of the Mn$^{2+}$ and W$^{6+}$ ions along the $a$-axis, both being octahedrally coordinated by oxygen\cite{Macavei93}. Along the $c$-direction the Mn sites carrying the partially frustrated $S=5/2$ spins form zig-zag chains. Cooling down from higher temperatures the spin-system orders at $T_{N3}=13.5$\,K into an incommensurate, collinear antiferromagnetic, sinusoidal spin-density wave with an easy spin-axis within the $ac$-plane. On further cooling at $T_{N2}\approx 12.6$~K a second order phase transition occurs into a non-collinear phase with the same propagation vector. In this phase a chiral spin-spiral with a spin-current not parallel to the propagation vector emerges breaking the spatial inversion symmetry. This results in the formation of electric polarization along the $b$-direction that can be well described by the inverse DM interaction \cite{Heyer06,Taniguchi06,Arkenbout06}. Finally, below $T_{N1}\approx 7.5$~K the spin arrangement becomes commensurate and collinear via a first-order phase transition, losing the ferroelectricity again.
To summarize with respect to the dielectric properties the multiferroic phase (ferroelectric with chiral spin structure) lies in the temperature range $T_{N1}=7.5$\,K\,$<T<T_{N2}=12.6$\,K, embedded between a paraelectric phase with collinear sinusoidal spin order at higher temperature and a paraelectric phase with collinear antiferromagnetic spin order at lower temperature (see Fig.~\ref{phases}).
In this article we report on broadband spectroscopic investigations of the linear and non-linear complex permittivity in high quality MnWO$_4$ single crystals above and within the multiferroic phase for frequencies from mHz to MHz and in electric fields up to 1000\,V/mm in order to shed light on the dynamical dielectric response of the system.

\subsection{Experimental details}

Single-crystals of MnWO$_4$ were grown from the melt using the top-seeded growth technique, as described in Ref.\,[\onlinecite{Becker07}]. The samples are ruby-red transparent and insulating above our measurement capabilities of 200~T$\Omega$ in the regarded temperature regime, which is indicative of a very low concentration of charge-doping defects. 
Structural and magnetic measurements confirmed the known behavior: The samples exhibit the monoclinic space group P2/c and show the above described sequence of phase transitions at $T_{N1}\approx 7.6$\,K, $T_{N2}\approx12.6$\,K, and $T_{N3}\approx13.5$\,K \cite{Heyer06,Taniguchi06,Arkenbout06}.
The dielectric measurements were performed in a commercial $^4$He-flow magneto-cryostat ({\sc Quantum-Design PPMS}) employing a home-made coaxial-line inset. The complex, frequency dependent dielectric response $\varepsilon^*(\nu)$ was measured using a frequency-response analyzer ({\sc Novocontrol}) for frequencies from 1\,mHz to 1\,MHz. 
For higher frequencies up to 200\,MHz a micro-strip setup was employed and the complex transmission coefficient ($S_{12}^*$) was evaluated via a vector network analyzer ({\sc Rohde \& Schwarz}). 
All  measurements were performed with the electric field along the crystallographic $b$-axis, the direction in which the spontaneous ferroelectric moment points in zero external magnetic field \cite{Taniguchi06,Arkenbout06}. If not denoted otherwise, the measurements of the complex permittivity were carried out with a stimulus of the order $E_{ac}\approx 1$\,V/mm. 
Additional measurements in higher fields up to 1000~V/mm were conducted employing a high-voltage option for the frequency response analyzer ({\sc Novocontrol HVB1000}). 
The non-linear permittivity contributions were obtained via the Fourier components of the dielectric response. These higher harmonics are directly evaluated from the normalized current response at multiples of the base frequency $n*\nu$ by the firmware of the frequency-response analyzer ({\sc Novocontrol}). 

The quasi static $P(E)$-measurements and the $P(T)$-data integrated from pyro-current measurements were recorded using a  high-precision electrometer ({\sc Keithley 6517B}).
In all cases, the contacts were applied to the plate-like single-crystals using silver paint in sandwich geometry with a typical electrode area of $A\approx 2$\,mm$^2$ and a sample thickness of $d\approx0.3$\,mm.
The uncertainty in the determination of the exact geometry together with additional (but constant) contributions of stray capacitances results in an uncertainty in the absolute values for the permittivity of up to 30\,\%.

\begin{figure}
\centerline{\includegraphics[width=0.85\columnwidth,angle=0]{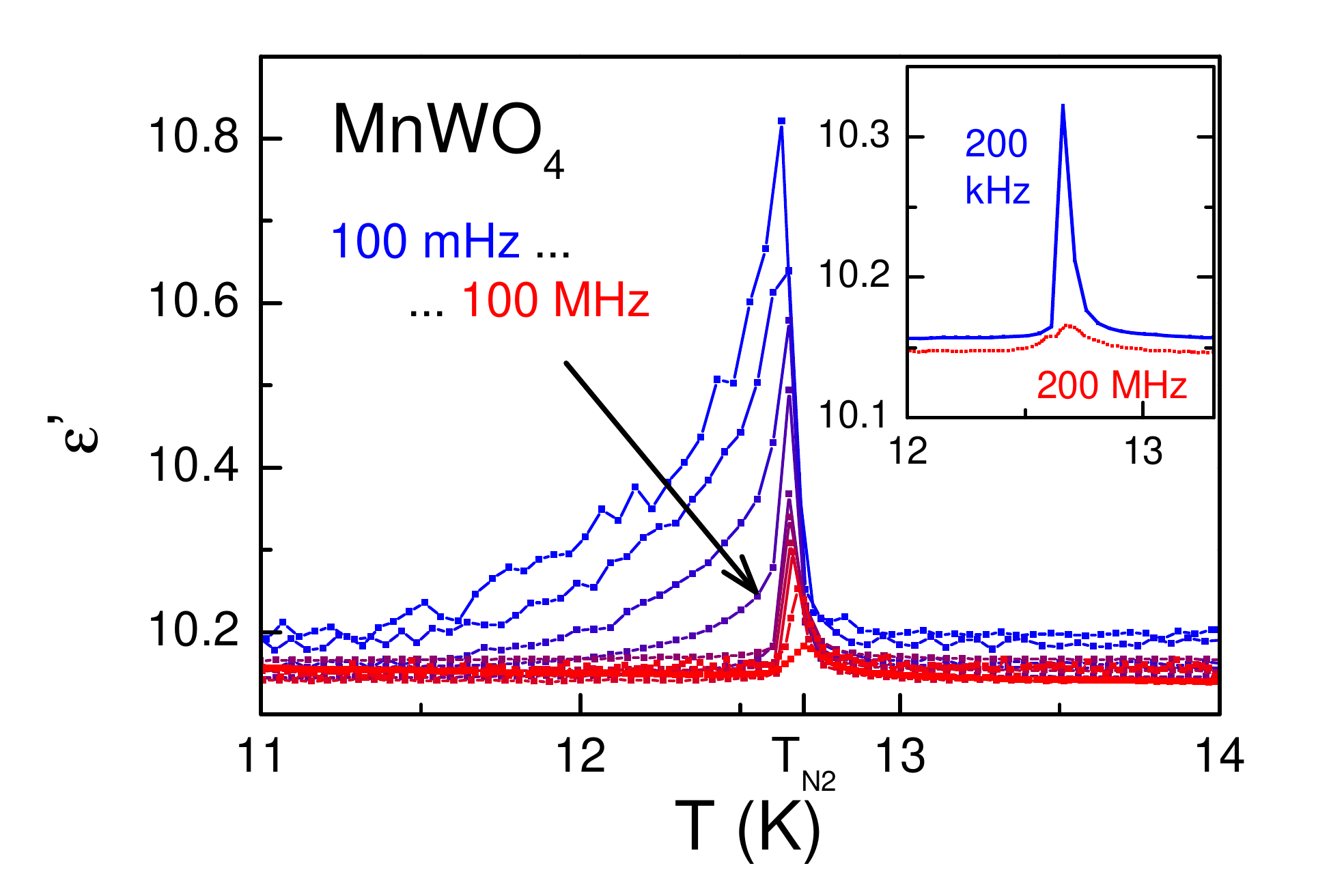}}
\caption{
(Color online) Real part of the dielectric permittivity $\varepsilon'$ of MnWO$_4$ for different frequencies between 100\,mHz and 100\,MHz in equidistant spacing measured with a stimulus of $\sim$1\,V/mm and as a function of temperature around the multiferroic transition at $T_{N2}\approx 12.6$\,K. The inset shows a zoom on higher frequencies.
}
\label{figeps1}
\end{figure}

In Fig.~\ref{figeps1} the real part of the dielectric permittivity $\varepsilon'(T)$ is shown for different frequencies covering a span of nine decades and a temperature region including both, the phase transition into a sinusoidal spin density wave at $T_{N3}\approx 13.5$\,K and the transition into the multiferroic phase at $T_{N2}\approx 12.6$\,K. While the paraelectric to paraelectric transition at $T_{N3}$ apparently has no pronounced effect on the linear dielectric permittivity, clear anomalies can be detected in the $\varepsilon'(T)$ curves at the paraelectric to multiferroic/ferroelectric transition at $T_{N2}$.
Characteristic for these anomalies is a distinguished frequency dependence. 
Usually permittivity measurements with such small stimuli are performed using capacitance bridges working at one single frequency, typically 1\,kHz or 10\,kHz. Using smaller frequencies broadens the tail in $\varepsilon'(T)$ below the transition. Also the peak height at the transition is strongly dispersive and suppressed for higher frequencies (see inset of Fig.~\ref{figeps1}).
The general behavior, i.e. having a flat non-dispersive dielectric background on the order of $\varepsilon_\infty\approx 10$, is in accordance with other published data \cite{Taniguchi06,Arkenbout06,Heyer06}.
The small weight of the anomaly at the transition compared to the dielectric background reflects, that the ferroelectric moment of the multiferroic phase is nearly four orders of magnitude smaller than in ordinary ferroelectrics where a Curie-Weiss-like divergence dominates $\varepsilon'(T)$ \cite{Blinc74}.
It has to be noted, that the small frequency dependent shift in $\varepsilon_\infty\approx 10$ may be due to 
apparatus-based uncertainties concerning the absolute values and shall not be regarded in the following, the relative uncertainty is much smaller of course.
In the following we discuss the dispersion 
within the multiferroic phase.

~

\subsection{Domain dynamics in the multiferroic phase}

\begin{figure}
\centerline{\includegraphics[width=0.8\columnwidth,angle=0]{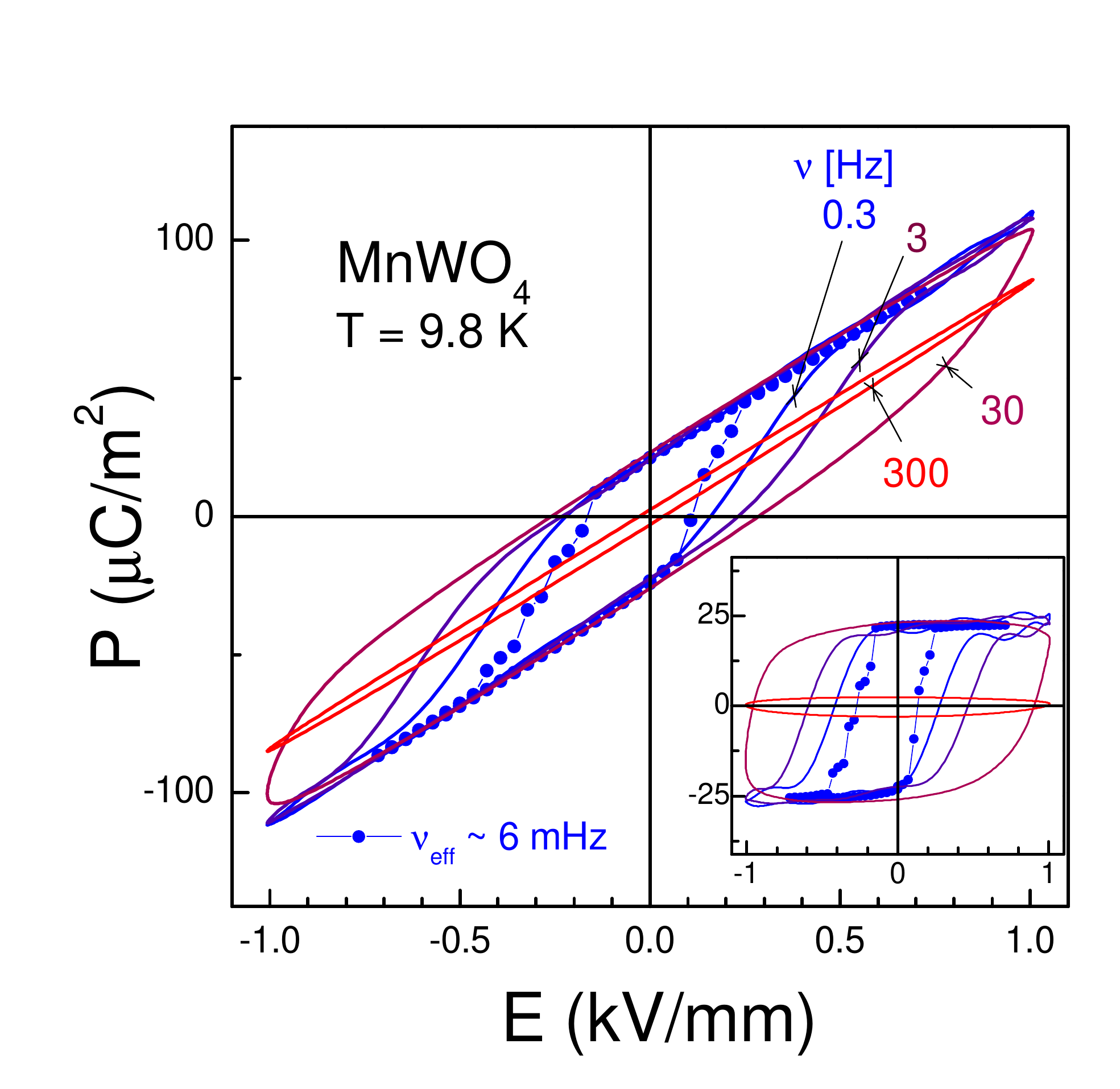}}
\caption{
(Color online) Electric field dependent polarization $P(E)$ measured for $E\parallel b$ at temperatures below $T_{N2}$. The dotted curve was measured using a  time dependent stimulus $E(t)$ with a triangular profile and an effective frequency of $\nu \approx 6$\,mHz. The other $P(E)$ loops were reconstructed from the  first 10~orders of the non-linear permittivity measured with sinusoidal stimulation at frequencies between 300~mHz and 300~Hz. The inset shows the $P(E)$ loops for the switchable net polarization after subtraction of the quasi-static dielectric background $\propto\varepsilon_\infty$.
}
\label{figPE}
\end{figure}

Below the transition into the multiferroic phase the dielectric response is dominated by ferroelectric domain contributions. Fig.~\ref{figPE} shows a hysteretic $P(E)$ curve measured along the crystallographic $b$-axis for $T\approx10$~K~$<T_{N2}$. The data (-$\bullet$-) are taken 'quasi-statically' using a time-dependent electric field with a triangular profile at an effective frequency of about 6~mHz. The other hysteresis loops displayed in Fig.~\ref{figPE} are measured with respect to sinusoidal stimulation at higher frequencies up to 300~Hz employing the higher-order Fourier coefficients of the non-linear permittivity $\varepsilon_n$ as discussed below \cite{Hemberger97}. 
The obvious frequency dependence of the polarization hysteresis will be discussed later, first we will address the general characteristics of the quasi-static behaviour.
An eye-catching feature of the $P(E)$ curve (-$\bullet$- in Fig.~\ref{figPE}) is the linear region extending up to high electric fields beyond the coercive field $E_c$. This slope corresponds to the dielectric background $\Delta P / \varepsilon_0\Delta E \approx \varepsilon_\infty\approx 10$, which is independent of the ferroelectric contributions within the multiferroic phase. As mentioned before, in these weakly ferroelectric multiferroic compounds the magnitude of this dielectric background contribution is comparable to the ferroelectric polarization itself. For this reason $P(E)$ cycles measured via the surface charge \cite{Kundys08} look different compared to other experiments: Finger {\em et al.} probed the switching of polarization indirectly via the coupling to the magnetic chirality via neutron diffraction in external electric fields\cite{Finger10}. Another method uses the nonlinear optical effect of second harmonic generation, as demonstrated in Refs.\,[\onlinecite{Meier09}] and [\onlinecite{Hoffmann11}]. For these latter experimental methods the dielectric background $\varepsilon_\infty$ does not contribute and the $P(E)$ loops directly exhibit a much more square-like shape. In order to visualize the characteristic parameters of the hysteresis curve, i.e.\ the coercive field $E_c$ at which the polarization jumps, one has to subtract this background, as displayed in the inset of Fig.~\ref{figPE}. The magnitude of the maximal jump in the polarization near the coercive fields corresponds to the switching of the sign of the complete spontaneous polarization  $\Delta P = 2P_s$. The value of $\Delta P\approx 50$\,$\mu$C/m$^2$ at $T=9.8$\,K is somewhat smaller than published results for the spontaneous polarization from pyro-current\cite{Arkenbout06,Taniguchi06} and $P(E)$ measurements \cite{Kundys08} but (considering the experimental uncertainty) still in rough agreement. 
However, already at this point we want to mention that the maximum of the switchable polarization $P_{sw}$ 
may differ from the spontaneous polarization $P_s$ due to local fields at defect sites, which may be frozen-in during pyro-current sequences. The dependency of such inner biases on prehistory and sample preparation will be discussed in a forthcoming publication \cite{Stein14}. In the following we tried to reduce all statically frozen-in electric fields by zero-field cooling and mainly used AC-fields oscillating symmetrically around zero. Nevertheless, we will in the following distinguish between the spontaneous and the switchable polarization. The latter in addition is a dynamic quantity. It can be seen directly from the inset of Fig.~\ref{figPE} that the effective value of $E_c$ is strongly frequency dependent. And if $E_c$ comes close to the amplitude of the stimulating electric field $E_{ac}$ also the value of the switched polarization $P_{sw}<P_{sw,0}$ is reduced and has to be distinguished from the maximal, quasi-statically switchable polarization.

\begin{figure}
\centerline{\includegraphics[width=0.8\columnwidth,angle=0]{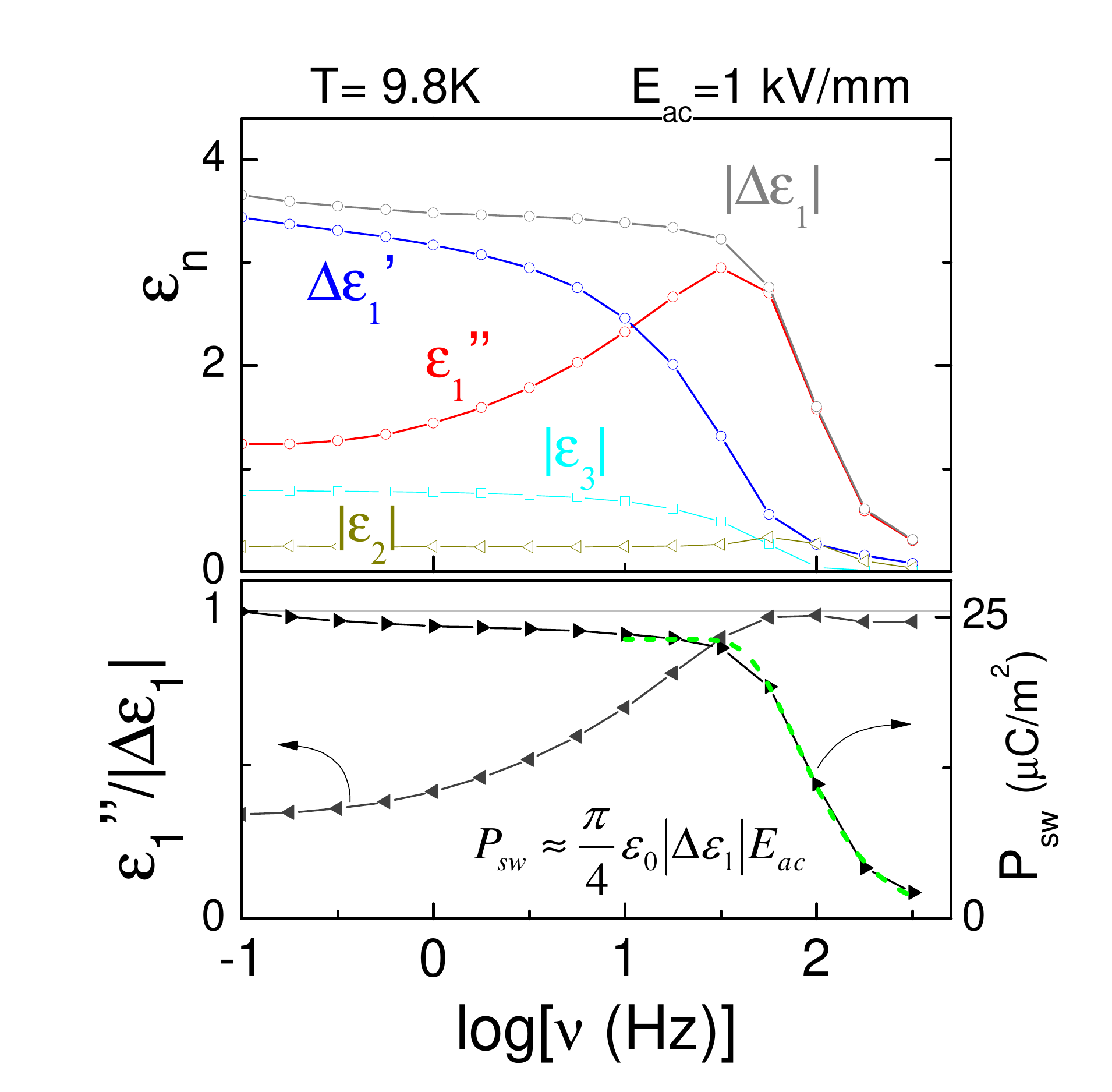}}
\caption{
(Color online) Upper frame: Spectra of the non-linear orders $\varepsilon_n$ of permittivity measured using a stimulus of $E_{ac}=1$\,kV/mm at $T=9.8$~K. The first-order component is displayed as $\Delta \varepsilon_1'= \varepsilon_1'-\varepsilon_\infty$ (blue), the dielectric loss $\varepsilon_1''$ (red) and the magnitude $|\Delta \varepsilon_1| = \sqrt{\Delta \varepsilon'^2 +\varepsilon''^2}$ (light grey). Also displayed is the magnitude of the second and third order permittivities $|\varepsilon_2|$ and $|\varepsilon_3|$ which denotes the non-linear character of the dielectric response within the multiferroic phase. 
Lower frame: The frequency dependence of the switched polarization $P_{sw}$ and the ratio $\varepsilon_1''/|\Delta\varepsilon_1|\approx E_c/E_{Eac}$.
The dashed line (green) represents a fit to $P_{sw}(\nu)$ following the Ishibashi-Orihara-model \cite{Ishibashi95} with a dimensionality parameter $d \approx 1.8$ as described in the text. 
}
\label{figeps123}
\end{figure}

In order to evaluate the dynamics of the ferroelectric switching behavior, i.e.\ the frequency dependence of the coercive field and the switched polarisation, it is favorable to take advantage of the Fourier coefficients $\varepsilon_n$ of the non-linear complex permittivity, as mentioned above ($n$ = order of Fourier coefficient).
The time dependent polarization response to a harmonic stimulation $E(t)=E_{ac}\sin\omega t$ can be expressed as 
\[
P(t)=\varepsilon_0E_{ac}\sum_{n=1}^{\infty}\left( \varepsilon_n'\sin n\omega t -\varepsilon_n''\cos n\omega t \right).
\]
The higher order components of the permittivity are experimentally determined via the higher harmonics of the frequency dependent polarization response using the lock-in technique.
To extract $E_c$ and $P_{sw}$ from the non-linear coefficients $\varepsilon_n$ recorded in the frequency domain gives the advantage of a much broader frequency range with much higher sensitivity compared to the direct recording of the equivalent time-domain response.
The higher sensitivity is caused mainly by the inherent lock-in concept used for the detection in the frequency domain measurement. 
In addition, for time-domain recording the shape of the hysteresis curves at higher frequencies would be strongly influenced by the measurement bandwidth. In the frequency domain 
the intrinsic dynamic values for $E_c$ and $P_{sw}$ can be deduced
already from the first order components, which will be discussed in the following.

Considering a ``perfect'', inversion-symmetric square-type hysteresis only the odd terms of $\varepsilon_n$ will contribute with a magnitude vanishing with $1/n$. 
To capture the exact shape of the loop one needs a considerable number of higher order terms. 
Fig.~\ref{figPE} displays $P(E)$-loops recalculated from the terms $\varepsilon_n$ with $n\leq 10$.
The dielectric background $\varepsilon_\infty$ only shows up in the real part of the first-order component $\varepsilon_1'$. Thus the purely ferroelectric contribution to the real part can be considered using the corrected first order contribution $\Delta\varepsilon_1'= \varepsilon_1'-\varepsilon_\infty$, the correspondingly corrected $P(E)$-loops are displayed in the inset of Fig.~\ref{figPE}. 
For a square-type hysteresis loop its height, i.e.\ the switchable polarization $P_{sw}$, is connected to the higher harmonics according to the Fourier representation of a step function. For the first order component it follows $\varepsilon_0E_{ac}|\Delta\varepsilon_1|=\frac{4}{\pi}P_{sw}$, where $|\Delta\varepsilon_1|=\sqrt{\Delta\varepsilon_1'^2+\varepsilon_1''^2}$ is the magnitude of the background corrected, i.e.\ only ferroelectric, contribution to the first order of the non-linear permittivity. Thus the switchable polarization can be derived already from the linear component\cite{Furukawa87}: $P_{sw}=\frac{\pi}{4}\varepsilon_0E_{ac}|\Delta\varepsilon_1|$. Furthermore, the phase of $\varepsilon_1^*$, is determined by the position $E_c$ within the $P(E)$ loop, at which the polarisation is switched. Usually, the imaginary part of the permittivity is associated with the dielectric loss. In the case of a (non-linear) hysteresis loop the energy per volume dissipated within one cycle $w_{cycle}$ is given by the inner area of the loop. For a square-type hysteresis loop this area can be estimated by the coercive field and the switchable polarisation via $w_{cycle}=2P_{sw} \cdot 2E_c$. Furthermore, the ratio between imaginary part and magnitude of the first order component define the ratio between the coercive field and the stimulating field amplitude: $E_c/E_{ac}=\varepsilon_1''/|\Delta\varepsilon_1|$. 
Using the above denoted expressions for $E_c$ and $P_{sw}$ one finds the familiar result that the dielectric loss is given by the imaginary part of the first order permittivity: $w_{cycle}=\pi\varepsilon_0\varepsilon_1''E_{ac}^2$. This result holds in general for arbitrarily shaped hysteresis loops driven with sinusoidal stimulus, only $\varepsilon_1''$ and not the higher-order terms contribute to the loss\cite{Furukawa87}. 
Coming back to the coercive field, the dielectric loss can be used in order to give an estimate: $E_c\approx E_{ac} {\varepsilon_1''}/{|\Delta\varepsilon_1|}$. This is of course restricted to the case $E_{ac} > E_c$, i.e.\ it is only valid as long as the coercive field does not exceed the stimulus. For ``real'' hysteresis loops, i.e.\ smeared out squares, this latter relation may not give directly the value of $E_c$ but at least the ratio $\varepsilon_1''/|\Delta\varepsilon_1|\approx 1$ may serve as an indication that the $P(E)$ loop is no longer completely saturated.

Anyway, for real $P(E)$ curves the coercive field has to be understood as an averaged quantity due to the distribution of switchable domains. The single-cycle quasi-static loop shown in Fig.~\ref{figPE} (-$\bullet$-) even shows Barkhausen jumps which demands an average over several cycles in order to determine an effective value for $E_c$. In addition, the simple zero-crossing of the polarization curve will depend not only on frequency but as well on the shape of the stimulating electric field: e.g., for a sinusoidal stimulus the switching will appear retarded compared to the a triangular excitation. Thus to capture the dynamical aspects of the polarization switching it is useful to define the effective frequency dependent coercive field via the maximum in the non-linear dielectric loss spectra $\varepsilon_1''(\nu)$, as outlined in the following.

In the upper frame of Fig.~\ref{figeps123} the permittivities $\varepsilon_{1,2,3}(\nu)$ are displayed for the temperature $T=9.8$~K, corresponding to the $P(E)$ data shown in Fig.~\ref{figPE}. 
Comparing $\varepsilon_{1}'$ and $|\varepsilon_{3}|$ for low frequencies where the polarisation can be fully switched we find roughly the factor 1/3 corresponding to the above mentioned expectation for square-like hysteresis loops. Also we find the even $|\varepsilon_{2}|$ to be much smaller as expected for symmetric hysteresis loops. But nevertheless $|\varepsilon_{2}|$ is still finite, denoting a slight asymmetry of the $P(E)$-loops due to the before mentioned possibility to freeze internal bias fields. However, due to the only small asymmetry we will not consider any bias effects in the following and regard the symmetrically averaged results.  

The main hysteresis features, i.e.\ the frequency dependent behavior of the coercive field and switchable polarisation, are characterized by the first order components of the non-linear permittivity  $\varepsilon_1'$ and $\varepsilon_1''$. 
With increasing frequency the dielectric loss $\varepsilon_{1}''(\nu)$ continuously rises until it reaches a well defined maximum and decays afterwards. This enhancement of $\varepsilon_{1}''(\nu)$ corresponds to an increased area within the $P(E)$ loops displayed in Fig.~\ref{figPE}. 
These dynamics of the ferroelectric polarisation switching are described considering a frequency dependent coercive field $E_c(\nu)$, which can be derived from the maxima in the dielectric loss spectra $\varepsilon_1''(\nu)$ at a given stimulus $E_{ac}$, i.e.\ for the frequency at which the coercive field $E_c(\nu)$ just reaches the external stimulus and the dissipative $P(E)$ area of the hysteresis loop becomes maximal. Up to this point the ratio $\varepsilon_1''/|\Delta\varepsilon_1|\approx E_c/E_{ac}$ (see lower frame of Fig.~\ref{figeps123}) can serve as an estimate for the coercive field.  
On increasing frequency the coercive field $E_c(\nu)$ reaches the stimulating field amplitude $E_{ac}$. This is denoted by the maximum $\varepsilon_{1}''(\nu)$ and the ratio $\varepsilon_1''/|\Delta\varepsilon_1|\approx 1$.  For higher frequencies only a further reduced fraction of the 
polarization can be switched within a field cycle, i.e.\ the $P(E)$ loops become flat and ellipsoid-like (see, e.g., the 300~Hz curve in the inset of Fig.~\ref{figPE}). The reduction of the switchable polarization $P_{sw}$ with increasing frequency due to the increase of the corresponding effective coercive field is directly monitored in the lower frame of Fig.~\ref{figeps123}. 
The steep drop down of $P_{sw}$ towards higher frequencies can be described using the Ishibashi-Orihara-model\cite{Ishibashi95,Orihara94} assuming growth limited domain dynamics. This model is based on switching dynamics dominated by the growth kinetics based on the Kolmogorov-Avrami model \cite{Avrami40}. 
Such a model represents the canonical approach to describe single crystalline materials, for which the switching is regarded as not 
being nucleation limited, in contrast to, e.g., polycrystalline films possessing small structural domains  \cite{Setter06}. 
In the present case it is adequate to assume quasi instantaneous electric-field stimulated nucleation of inverse domains and their subsequent growth in $d$ dimensions is considered. 
The dashed green line in Fig.~\ref{figeps123} was fitted using the expression $P_{sw}/P_{sw,0} = 1 - \exp(-f^{-d}\Phi_E)$, where the exponent $d\approx 1.8$ denotes the effective growth dimension of the domains and $\Phi_E$ is a factor depending on the stimulating field amplitude.\cite{Ishibashi95}   
This decay of the polarisation switching is relatively narrow. The right flank of the loss peak width, from the maximum to the half maximum $\log(\Delta \nu_\text{HWHM}/\nu_0) \approx 0.5$, is even narrower than the corresponding width for the mono-dispersive Debye-relaxator of 0.57 decades. This may point towards the absence of a wide distribution of domain sizes, which correspondingly would offer a broadened distribution of relaxation or switching times. Mono-dispersive, Debye-like permittivity spectra were also reported for the small field response of domains at the multiferroic transition in rare earth manganites \cite{Schrettle09,Kagawa09}. But, as indicated above, the shape of the loss spectrum in the upper frame of Fig.~\ref{figeps123} has rather to be explained by the assumption of a scenario of growth limited polarisation processes than by a static distribution of relaxation times. The very broad left slope of the loss peak for lowest frequencies may be determined by pinning, 
followed by a regime where the coercive field, which is determined by the domain growth rate, effectively increases with increasing stimulating frequency. From the decay of the switched polarisation an effective growth dimensionality of $d \approx 1.8$ can be estimated. Even though the dimensional meaning of this number should not be overemphasized it is in accord with the reduced dimensionality observed via the direct SHG-imaging of the domains, which have the shape of elongated bubbles and show one-dimensional growth \cite{Meier09}. A small orientational preference with respect to the magnetic easy-axis and a shape dependence on the electric poling history was observed\cite{Meier09, Hoffmann11}. The domains are large objects with extension above $\mu$m and for the preferred direction even up to mm, i.e.\ mesoscopic objects with a preferred growth axis. 

\begin{figure}
\centerline{\includegraphics[width=1.0\columnwidth,angle=0]{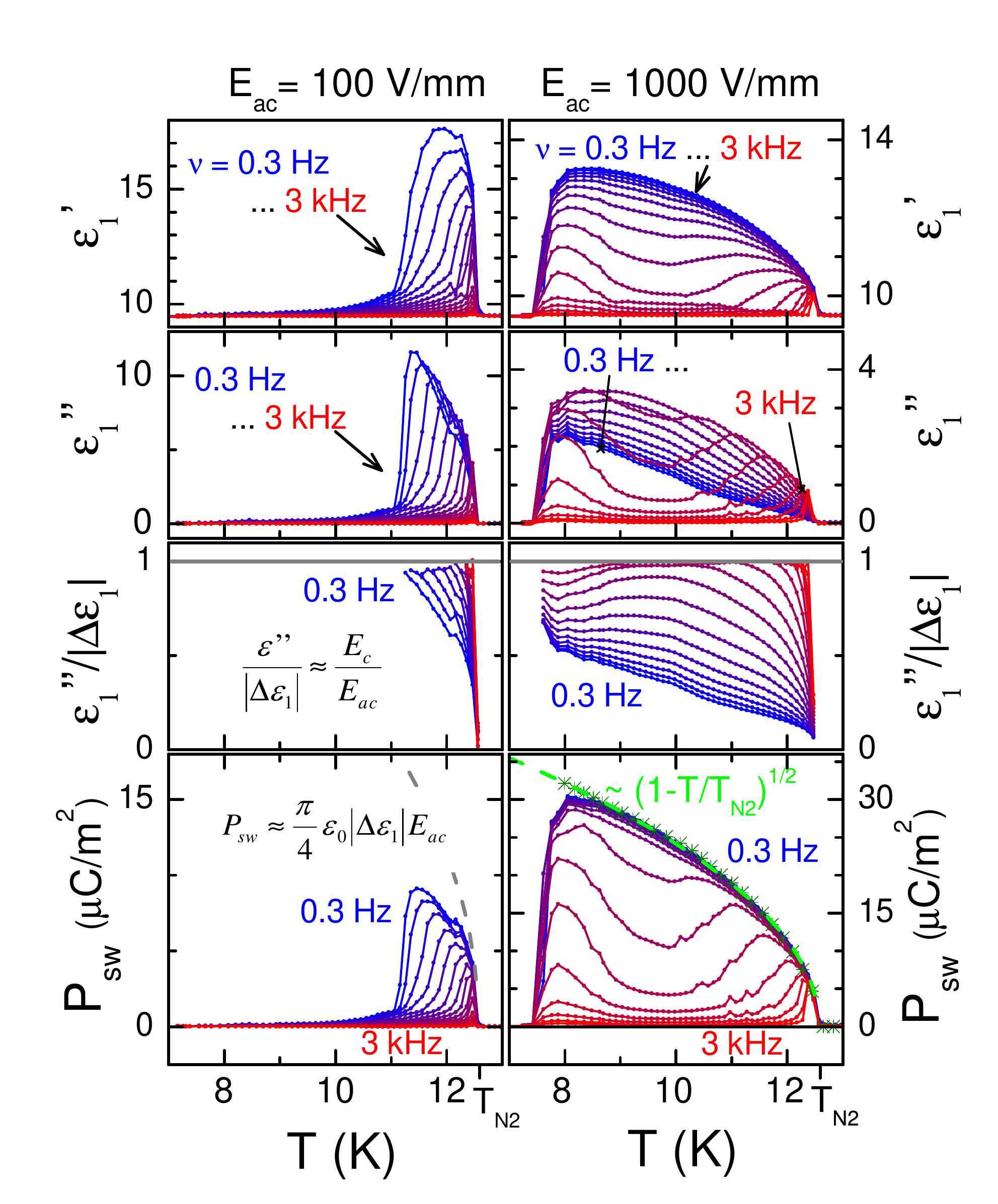}}
\caption{
(Color online) Real and imaginary part of the first-order component of the non-linear permittivity measured for frequencies between 0.3\,Hz and 3\,kHz at two different stimuli of 100 (left column) and 1000\,V/mm (right column). 
The additional data shown in the bottom right frame (*) originates from integrated pyro-current measurements and traces the temperature dependence of the spontaneous polarization. 
The dashed line is a fit according to a mean-field like square-root behavior.
}
\label{figPsvonT}
\end{figure}

In order to gain information on the temperature dependence of the switchable polarization and the dynamic coercivity within the multiferroic phase of MnWO$_4$ we measured the complex first-order non-linear permittivity for frequencies between 0.3 and 300~Hz for driving fields $E_{ac}=100$ (left column of Fig.~\ref{figPsvonT}) and 1000~V/mm (right column of Fig.~\ref{figPsvonT}). 
The upper frames of Fig.~\ref{figPsvonT} display the real and imaginary part of the first-order permittivity $\varepsilon_{1}$, which show a strong dependency on frequency $\nu$ and stimulus $E_{ac}$. For the lower frequencies and the higher stimulus a contribution to both, $\varepsilon_{1}'$ and $\varepsilon_{1}''$, can be detected throughout the complete multiferroic/ferroelectric temperature regime $T_{N1}\leq T\leq T_{N2}$. This is not the case for high $\nu$ or low $E_{ac}$: here a contribution appears below $T_{N2}$ but then vanishes towards lower temperatures already within the ferroelectric phase above $T_{N1}$. For the data taken at 3~kHz and 100~V/mm only a very narrow peak can be seen comparable to the usual anomalies reported in the literature \cite{Arkenbout06,Taniguchi06} 
or in Fig.~\ref{figeps1}.
The use of a stimulation of 100~V/mm, which does not saturate the hysteresis loops throughout the entire multiferroic temperature regime, bridges the dielectric results gained for small and high stimulus measurements. Even for small stimulus there is a temperature regime for which the driving field indeed does saturate the loops due to the small enough coercive field. If the coercive field grows beyond the stimulus, the contribution to $\varepsilon_1$ vanishes. The anomaly observed in $\varepsilon_{1}'(T)$ is narrower if the employed driving field is smaller.  

The underlying scenario can 
be visualized by the switchable polarization $P_{sw}$ and the estimation for the coercive field $E_c/E_{ac}\approx\varepsilon_1''/|\Delta\varepsilon_1|$ in the lower frames of Fig.~\ref{figPsvonT}. The coercive field rises on cooling below $T_{N2}$. When $\varepsilon_1''/|\Delta\varepsilon_1|$ comes close to unity, i.e.\ when the temperature- and frequency-dependent coercive field reaches the amplitude of the applied stimulus, the corresponding switched polarization $P_{sw}$ is drastically reduced. Only in the quasi-static limit, for high stimulus and low frequencies, a large polarization $P_{sw,0}$ can be switched throughout the whole multiferroic temperature regime. In the absence of frozen-in bias internal fields 
one can assume the quasi-statically polarization to trace the full spontaneous polarization, i.e.\ $P_{sw,0} \propto P_s$. For this case ($E_c<E_{ac}$) the $\varepsilon_1$ data scale and reveal the temperature dependence of the spontaneous polarization (see bottom right frame of Fig.~\ref{figPsvonT}). This temperature dependence is in agreement with results from pyro-current measurements 
(displayed in the bottom right frame of Fig.~\ref{figPsvonT}) 
and can be described 
by the mean field expression for the continuous phase transition $P_s\propto(\frac{T_{N2}-T}{T_{N2}})^\beta$ with $\beta=1/2$, which is derived from a Landau-type approach with respect to symmetry considerations  \cite{Toledano10} (see dashed line in the bottom frames of Fig.~\ref{figPsvonT}). 
Indeed the ferroelectric polarization in MnWO$_4$ was shown to scale with the product of the two magnetic amplitudes forming the chiral magnetic structure below $T_{N2}$ \cite{Finger10}.
Towards the lower phase boundary at $T_{N1}$ the switchable polarization drops abruptly, as expected for a first-order transition. 

\begin{figure}
\centerline{\includegraphics[width=0.9\columnwidth,angle=0]{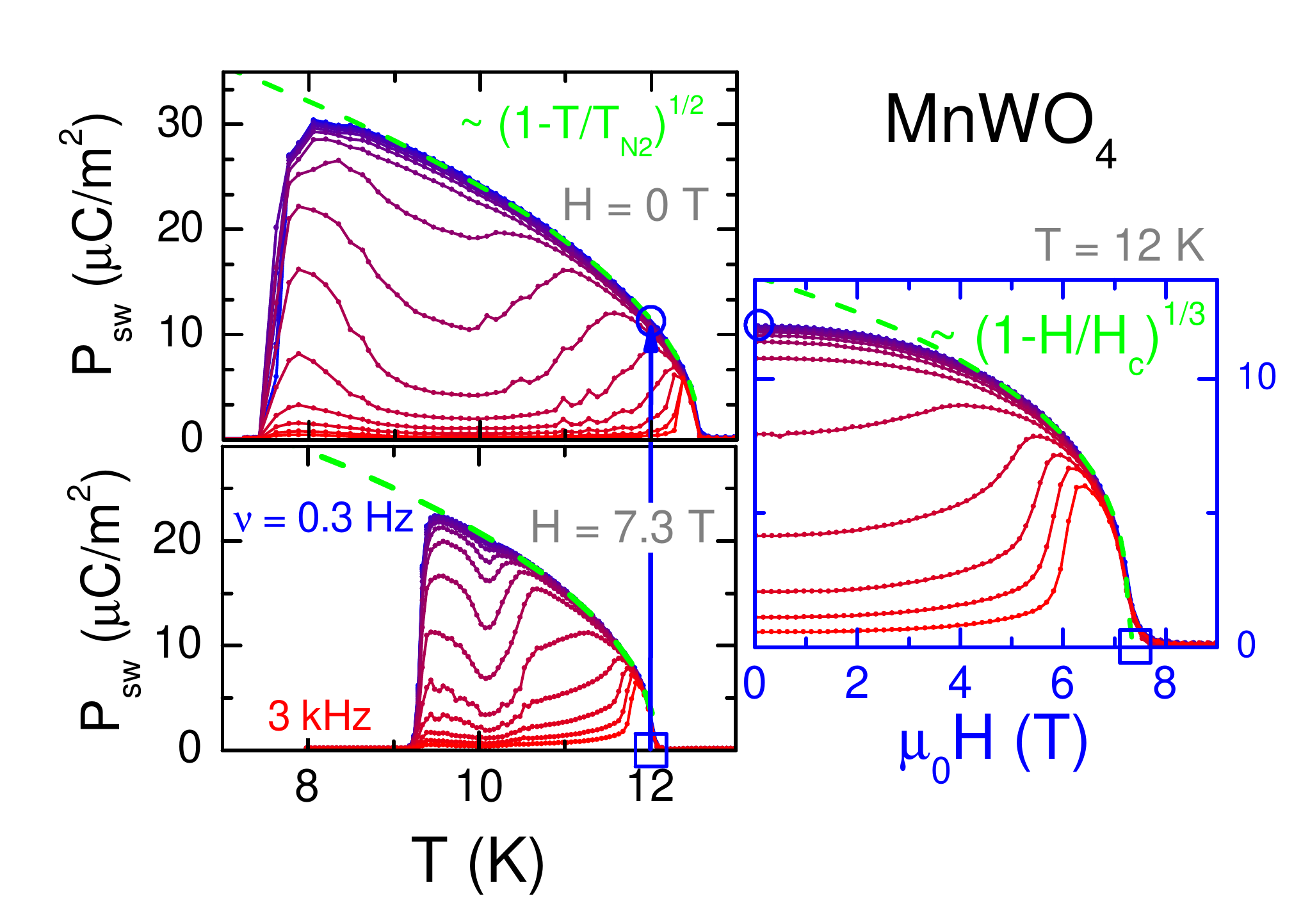}}
\caption{
(Color online) Switchable polarization $P_{sw}$ 
measured for frequencies between 0.3\,Hz and 3\,kHz at a stimulus of 1\,kV/mm.
Left frames: Temperature dependence at zero field and $\mu_0H=7.3$\,T. Right frame: Magnetic field dependence at $T=12$\,K.
}
\label{figPsvonB}
\end{figure}

The observed switching behavior is not significantly altered applying an external magnetic field. The corresponding data are displayed in Fig.~\ref{figPsvonB} together with the zero field results (already shown in the bottom right frame of Fig.~\ref{figPsvonT}). Regarding the temperature dependent data in an external magnetic field of $B=7.3$\,T it can be noticed that the multiferroic temperature regime is narrowed, i.e.\ $T_{N2}$ is lowered and $T_{N1}$ is increased by the external magnetic field in accordance with the published $(B,T)$-phase diagrams \cite{Arkenbout06,Taniguchi06}. Again, the square-root behavior for the rise of the spontaneous polarization below $T_{N2}$ can be deduced from the quasi-static limit of the switchable polarization. A similar mean-field scaling holds as well for the field dependence of the polarization displayed in the right frame of Fig.~\ref{figPsvonB}. At a temperature of $T=12$\,K\,$<T_{N2}$ one can continuously drive the spontaneous polarization to zero, which can be interpreted as the quasi-static limit of the frequency dependent value for the switchable polarization $P_s=P_{sw,0}$. On approaching the critical field, in this case $H_c\approx 7.3$\,T, the polarization vanishes as $P_s\propto(\frac{H_{c}-H}{H_{c}})^{\beta_H}$ with $\beta_H=1/3$, which might be expected, according to canonical Landau-theory, for a ferroelectric with a linear magnetoelectric contribution to the free energy near the multiferroic phase transition \cite{Kim09,Toledano10}. One has to note that in the present case the spontaneous polarization is only a secondary order parameter. 


\begin{figure}
\centerline{\includegraphics[width=0.9\columnwidth,angle=0]{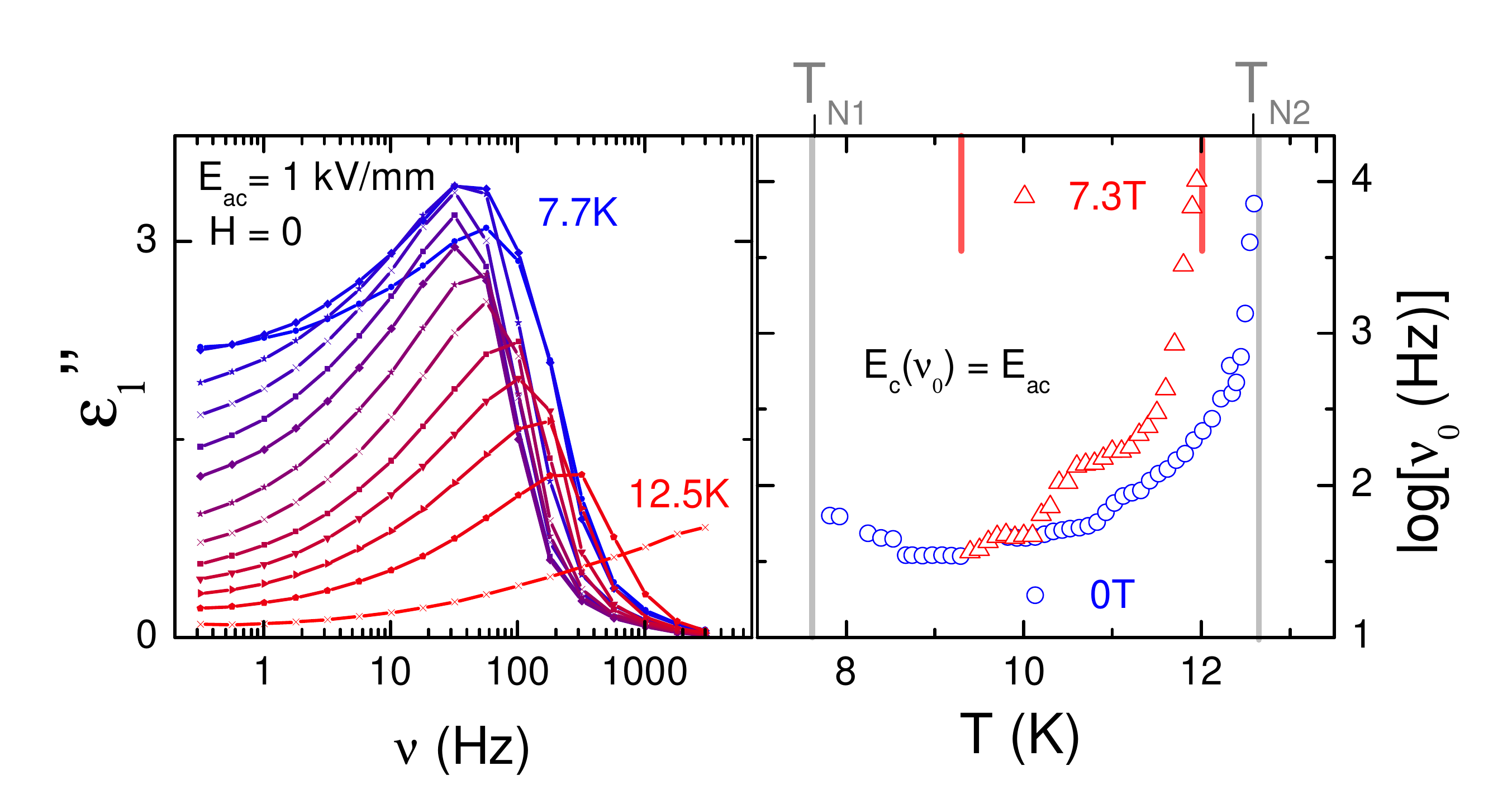}}
\caption{ 
(Color online)  Dielectric loss spectra (left frame) measured for a stimulus of 1000\,V/mm. From the maxima in these spectra a temperature dependent mean switching time can be evaluated (right frame). The right frame also displays the corresponding results for an applied magnetic field of 7.3~T (red symbols).
}
\label{figf0}
\end{figure}

Turning back to the polarization dynamics below $T_{N2}$ one can see directly from Fig.~\ref{figPsvonB} that the switching process is getting faster in the vicinity of the second order phase transition, no matter whether driven by temperature or magnetic field. For the temperature dependent data one has to note the minimum in the $P_{sw}(T)$-curves for the higher frequencies around $T\approx 10$\,K, which also persists in external magnetic field, as demonstrated via the $P_{sw}(T,H=7.3\mathrm{\,T})$-curves. Such a minimum also was found in dielectric measurements by Kundys {\em et al.} \cite{Kundys08} and in polarized neutron measurements sensing the switchability of the magnetic chirality by an electric field \cite{Baum13}. This minimum denotes that the effective coercive field for a given frequency slightly re-decreases on approaching $T_{N1}$ and the switching gets 'softer' again for temperatures below 10~K. Actually, such a re-acceleration of the domain dynamics is unexpected as the phase transition at $T_{N1}$ is discontinuous. However, in contrast to the continuous phase transition at $T_{N2}$ the coercive field for a given frequency and the effective switching time for a fixed stimulus stay finite on approaching the first-order transition at $T_{N1}$ from above. 

The softening of the domain switching can 
be monitored via the effective switching frequency $\nu_0$, as gained from the maxima in the dielectric loss $\varepsilon_1''(\nu)$ (left frame of Fig.~\ref{figf0}). 
The loss spectra displayed in Fig.~\ref{figf0} cover the complete multiferroic temperature regime but resemble the pronounced asymmetry of the exemplary loss data of Fig.~\ref{figeps123} recorded at 9.8~K. As already discussed the shape of these loss spectra reflects the time and accordingly frequency dependence of the switching process. These dynamics include the field-driven nucleation and growth of domains, which can effectively be described via a frequency dependent coercive field $E_c(\nu,T)$ together with slower processes, such as the creep of domain walls \cite{Ishibashi95,Kleemann07}. Thus the shape of the spectra $\varepsilon_1''(\nu)$ should not be related to an asymmetric distribution of relaxation times for which the position of the maximum not necessarily gives the mean relaxation rate. The maxima in the present case rather denote the maximal dissipation, which can be expected for the frequency dependent coercive field reaching the stimulus. This defines an effective switching rate for a given stimulus, as it is displayed in the right frame in Fig.~\ref{figf0}. (Of course one also could interpret the data in terms of an effective switching time $\tau_0=1/2\pi\nu_0$.) 
The steep increase of $\nu_0$ in the vicinity of $T_{N2}$ denotes the softening of the switching process in the vicinity of the continuous multiferroic phase transition.
However, away from the second order multiferroic phase transition, the switching time for stimulating fields of 1~kV/mm lies in the millisecond range, which is relatively slow compared to canonical ferroelectrics (e.g., \cite{Scott96,Setter06}), but is in accordance with reports on such slow switching in MnWO$_4$ obtained in time domain studies \cite{Baum13,Hoffmann11}. The effective switching rate forms a minimum and slightly re-increases on approaching $T_{N1}$, as already implied by the $P_{sw}(T)$ data discussed above. And again, despite the slight, unexpected re-acceleration of the domain dynamics on further cooling the effective switching time for a fixed stimulus stays finite on approaching the first-order transition at $T_{N1}$ in contrast to the divergent behavior close to the continuous phase transition at $T_{N2}$.

\begin{figure}
\centerline{\includegraphics[width=1\columnwidth,angle=0]{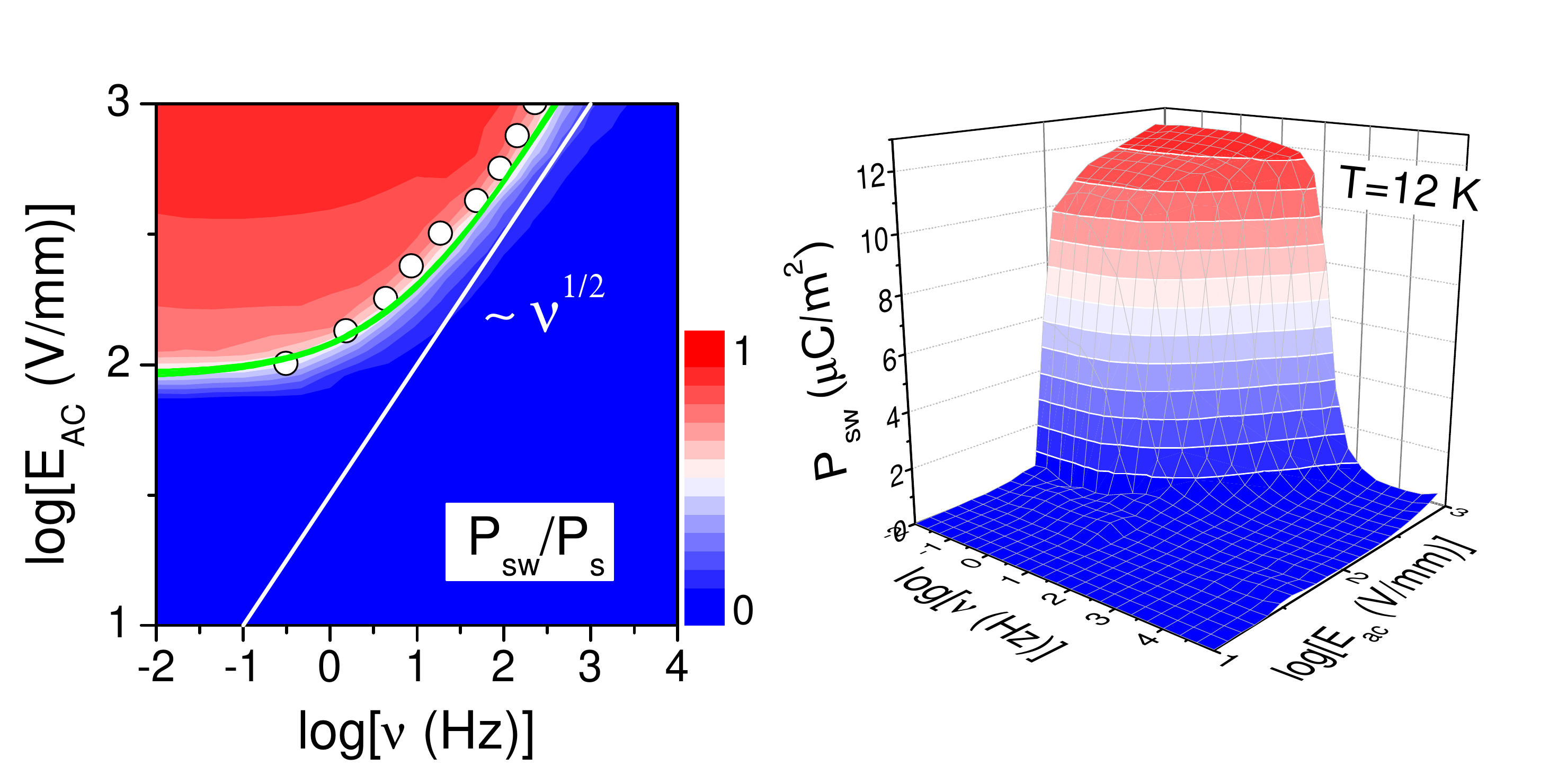}}
\caption{
(Color online) Contour plot of the switchability $\Delta|\varepsilon| E_{ac} / P_S$. The solid line is a fit according to the modified Ishibashi-Orihara model\cite{Ishibashi95} for the frequency dependence of the coercive field as described in the text. The symbols ($\circ$) show the position of the maxima in the dielectric loss spectra for the corresponding stimulus.
}
\label{figcolorplot12K}
\end{figure}

To elucidate the frequency dependence of the effective coercive field we measured the non-linear dielectric response at a fixed temperature of $T=12$\,K within the multiferroic phase for different frequencies
100~mHz\,$\le\nu\le$\,10~kHz and different electric stimuli $E_{ac}$ between 10 and 1000~V/mm. The results for the switchable polarization $P_{sw}$ are displayed in the right part of Fig.~\ref{figcolorplot12K}. For low frequencies and high stimuli nearly the complete spontaneous polarization, which at this temperature is about $P_s\approx 12$\,$\mu$C/m$^2$, can be switched; for high frequencies or small stimuli nearly no polarization can be switched. A quantity which characterizes the {\em switchability} of the multiferroic domains can be defined as the effective switchable polarization normalized to the maximal switchable, i.e.\ the spontaneous,
polarization $P_s$. The switchability is displayed in the left frame of Fig.~\ref{figcolorplot12K} as a colour-coded contour plot ranging between unity (red, fully switchable) and zero (blue, not switchable). The symbols ($\circ$) in Fig.~\ref{figcolorplot12K} denote the position $\nu_0(E_{ac}\approx E_c)$ of the maxima in the dielectric loss $\varepsilon''(\nu, E_{ac})$. These points are systematically shifted to lower frequencies by roughly a quarter of a decade compared to the value $P_{sw}/P_s=1/2$ (color coded white); for a comparison of the position of the loss maxima to the position of $\Delta|\varepsilon_1|=1/2$ see also Fig.~\ref{figeps123}. Both curves can be used to quantitatively describe the frequency of an effective coercive field. The green solid line in the left part of Fig.~\ref{figcolorplot12K} is again fitted according to the Ishibashi-Orihara model \cite{Orihara94,Ishibashi95} 
which predicts a frequency dependence $E_c\propto \nu^{d/\alpha}$. 
Here $d$ is again the growth dimension of the domains and $\alpha$ a constant determining the field dependence of the factor $\Phi_E\propto E^\alpha$ defined above \cite{Ishibashi95}. 
Together with a frequency independent offset due to pinning effects we get an expression $E_c \propto E_{pin}+\nu^{d/\alpha}$. The value for the pinning field lies around $E_{pin}\approx 90$\,V/mm, which sets a lower boundary for the effective coercive field: For smaller stimulating fields no switching occurs even for lowest frequencies. The exponent of the frequency dependence of $E_c(\nu)$ can be read off as the slope of the $E_c(\nu)$ curve in the double logarithmic representation of Fig.~\ref{figcolorplot12K} to be  $d/\alpha\approx 0.5$, which together with the value of $d\approx 1.8$ for the effective growth dimension 
suggest a value of $\alpha\approx 3.6$. This value is considerably smaller compared to findings in proper ferroelectrics, where a value of $\alpha\approx 6$ was found \cite{Ishibashi95,Scott96} revealing a smaller influence of the driving field on the effective switching time.    

A reason for the influence of pinning and the slow switching process may be connected to the very small electric dipole moment, which lies about three orders of magnitude below the ordered moment in proper ferroelectrics. The domain wall widths stay broader than in typical ferroelectrics, as expected for magnetically driven domain structures. Usually pinning of magnetic domain walls due to impurities is much less pronounced than the pinning of electric domain walls due to defect induced electric random fields.
In this magnetoelectric case pinning by electric random fields can occur and especially for such spiral spin structures a small magnetic anisotropy may lead to bulk pinning effects \cite{Nattermann12}.
Furthermore, the magnetic modulation is coupled with a $2\times q_\textrm{magnetic}$ structural modulations, which show coherent interferences \cite{Finger10b}.  
A consequence of the low domain wall mobility and the high value for ${d/\alpha}\approx 1/2$ is the high coercive field, which can be extrapolated for the high-frequency case and might be relevant with respect to application: The estimated value for $E_c(1~\mathrm{GHz})\approx $~1\,MV/mm is clearly beyond the typical values for the electric break-through. However, this obstacle - asides from the general problems of frustration related low transition temperatures and residual conductivity - might be overcome by the reduction of the growth dimension and domain volume using MnWO$_4$ as thin films.

\subsection{Summary}

In MnWO$_4$ the coupled magnetoelectric dynamics can be studied using broadband linear and non-linear dielectric spectroscopy for $E||b$, the crystallographic axis along which spontaneous polarization forms below $T_{N2}$. 
Within the ordered multiferroic regime the dielectric response is determined by the switching of ferroelectric domains and the growth velocity of domain walls, respectively. The quasi-static limit of the switchable polarization $P_{sw}$, which traces the spontaneous polarization $P_{s}$, exhibits mean-field behavior with a critical exponent $\beta =1/2$ for the temperature-driven transition and $\beta_H =1/3$ for the magnetic-field-driven transition near $T_{N2}$. The mean relaxation rate reaches values in the 10~mHz range in the middle of the multiferroic temperature regime and exhibits an unexpected  re-acceleration towards the first-order phase transition at $T_{N1}$. The electric field dependence of the polarization-switching dynamics can be described by a frequency-dependent coercive field $E_c \propto E_{pin}+\nu^{d/\alpha}$ using a considerable pinning threshold $E_{pin}$. 
The value for the frequency exponent is roughly $d/\alpha \approx 1/2$ with an effective growth dimension of $d\approx 1.8$, considering a growth-limited scenario as described via the Ishibashi-Orihara model.
It remains an open question whether the unusually slow domain dynamics in this compound can be modified via a further reduction of the effective dimension, e.g., in thin films. Also the details of the strong pinning effect and its possible alteration, e.g., via static poling or controlled embedding of defects needs further studies. 

We demonstrated that the dielectric response of MnWO$_4$ comprises a very broadband dynamics 
ranging down to mHz frequencies for the domain switching processes in the middle of the multiferroic phase. It will be interesting to compare these dynamical characteristic to multiferroics possessing different magnetoelectric coupling mechanisms or a simpler sequence of magnetic transitions.

{\bf Acknowledgements}

This work has been funded by the DFG through SFB608 (Cologne) and Institutional Strategy of the University of Cologne within the German Excellence Initiative.


\end{document}